# The forgotten pillar of sustainability: development of the S-assessment tool to evaluate Organizational Social Sustainability

**Alessandro Annarelli, Tiziana Catarci, Laura Palagi**

DIAG - Department of Computer, Control, and Management Engineering Antonio Ruberti

**Abstract**

Pursuing sustainable development has become a global imperative, underscored adopting of the 2030 Agenda for Sustainable Development and its 17 Sustainable Development Goals (SDG). At the heart of this agenda lies the recognition of social sustainability as a pivotal component, emphasizing the need for inclusive societies where every individual can thrive. Despite its significance, social sustainability remains a "forgotten pillar," often overshadowed by environmental concerns. In response, this paper presents the development and validation of the S-Assessment Tool for Social Sustainability, a comprehensive questionnaire designed to evaluate organizations' performance across critical dimensions such as health and wellness, gender equality, decent work, and economic growth, reducing inequalities, and responsible production and consumption. The questionnaire was constructed on the critical dimensions identified through a systematic and narrative hybrid approach to the analysis of peer-reviewed literature. The framework has been structured around the values of the SDGs. It aims to empower organizations to better understand and address their social impact, fostering positive change and contributing to the collective effort towards a more equitable and sustainable future. Through collaborative partnerships and rigorous methodology, this research underscores the importance of integrating social sustainability into organizational practices and decision-making processes, ultimately advancing the broader agenda of sustainable development.

**Introduction**

In 2015, the global community took a monumental step forward in the pursuit of a sustainable and equitable world by adopting the 2030 Agenda for Sustainable Development. This landmark agreement, endorsed by all United Nations Member States, laid out 17 ambitious Sustainable Development Goals (SDGs) aimed at addressing pressing issues such as poverty, inequality, and environmental degradation. These goals serve as a comprehensive roadmap for collective action, guiding nations, organizations, and individuals towards a future where prosperity is shared, justice prevails, and our planet thrives.

At the heart of this agenda lies the recognition that sustainability is not merely an environmental concern but a multifaceted concept encompassing social, economic, and environmental dimensions (Fox and Stallworth, 2015). While efforts to mitigate climate change and protect biodiversity are crucial, true sustainability also demands attention to the intricate social dynamics shaping our world (Abu-Ghaida and Klasen, 2004). Within this context that the concept of Social Sustainability emerges as a pivotal component of the broader sustainability discourse (Fujimoto and Härtel, 2017).

Social Sustainability, often referred to as the "forgotten pillar" of sustainability, emphasizes the importance of fostering inclusive societies where every individual can thrive, regardless of their background, identity, or circumstances (Fischer et al., 1993). It encompasses a wide array of interconnected issues, including but not limited to human rights, social justice, gender equality, diversity, and mental health (Gaard, 2001; Acker, 2012; Boimabeau, 2009; Boudiny, 2013). Recognizing and addressing these challenges is not only a moral imperative but also a strategic necessity for organizations operating in today's complex and interconnected global landscape (Berman and Paradies, 2010; Fuller and Young, 2022).

As we embark on this journey towards a more sustainable future, it is imperative that businesses play a central role in driving progress towards the SDGs. With their vast resources, influence, and reach, businesses have the potential to catalyze meaningful change and accelerate the transition towards a more just and sustainable world. However, to effectively fulfill this role, businesses must first understand their impact on society and the environment, and this requires robust frameworks for assessment and measurement (Flammer and Luo, 2017).

It Is within this context that the research project takes shape. Partnering with ReWorld[1] and ESG Culture Lab (Eikon Strategic Consulting[2]), we set out to develop a comprehensive questionnaire aimed at evaluating organization'' performance in the realm of Social Sustainability. Drawing inspiration from the 16 points of the ReWriters manifesto[3], that mimics the SGDs, our questionnaire seeks to assess organizations across a range of critical dimensions, including intergenerational justice, diversity, inclusivity, and mental health.

Through this endeavor, we aim to empower organizations to better understand and address their social impact, thereby contributing to the collective effort towards achieving the SDGs. By providing a practical tool for organizations to measure and improve their performance in the social sustainability domain, we hope to catalyze positive change and pave the way for a more equitable and sustainable future for all.

In the following sections of this paper, we will provide a detailed description of our research methodology and present the questionnaire developed as a result of our collaborative efforts with ReWorld.

---

[1] https://re-world.it/
[2] https://www.eikonsc.com/
[3] https://rewriters.it/manifesto/

**Methodology**

From a methodological point of view, a literature review was conducted with a hybrid approach. Indeed, we employed typical elements of the systematic approach for the research and selection of the papers, and of the narrative approach for the subsequent analysis. For each point and sub-point of the ReWriters's manifesto, a specific research key has been defined, which may involve logical operators. The aim, at least in the initial phase of the work, to conduct a separate literature review for each relevant element. The literature search was conducted on Scopus[4] by using a systematic approach based on keywords search.

For each search key, the articles were filtered according to the following criteria:

- Subject areas: (1) Business, Management and Accounting; (2) Social Sciences; (3) Economics, Econometrics and Finance; (4) Decision Sciences
- Source: Journal
- Language: English

The intent was, therefore, to prefer articles published in peer- reviewed scientific journals, for reasons mainly of scientific rigor.

Subsequently, a narrative approach was followed, and a further selection was made based on analysis of the title and abstract of the articles belonging to the first quartile in terms of a number of citations received, favoring the most recent articles, where possible.

The result of the search and selection is reported in Table 1 where for each SDGs' topic (TOPIC), the used keyword (KEYWORD), the overall number (ALL) of papers retrieved and of the number of the selected ones (FILTERS) are reported.

Given the variety of topics covered by each point of the above-referenced ReWriters manifesto, we decided to de-structure those in multiple topics, so as to ensure a more complete and thorough analysis.

Table 1: key dimensions of analysis with corresponding search words

| TOPIC | KEYWORD | ALL | FILTERS |
|---|---|---|---|
| BODY POSITIVITY | body AND positivity | 4880 | 146 |
| WORKLIFE BALANCE | work-life AND balance | 9332 | 3727 |
| DIGITAL RESPONSIBILITY | digital AND responsibility | 6315 | 1902 |
| INCLUSIVE LANGUAGES | inclusive AND language | 5898 | 2568 |
| NON-VIOLENT COMMUNICATION | nonviolent AND communication | 216 | 96 |

---

[4] https://www.scopus.com/

| TOPIC | KEYWORD | ALL | FILTERS |
|---|---|---|---|
| MEDIA ETHICS | "media ethic*" OR "journal* ethic*" | 1209 | 620 |
| COLLECTIVE INTELLIGENCE | "social intelligence" OR "collective intelligence" | 5880 | 1173 |
| COOPERATION AS THE KEY TO DEVELOPMENT | cooperation AND development | 104620 | 25527 |
| AFFECTIVE EDUCATION | "emotional education" | 439 | 127 |
| MENTAL HEALTH | "mental health" | 456031 | 66537 |
| QUEERNESS | queerness | 1504 | 772 |
| LGBTQI+ RIGHTS | lgbt AND rights | 1837 | 1053 |
| INTERSECTIONALITY | intersectionality | 12444 | 7332 |
| CONTRAST TO ABLEISM | ableism | 1223 | 707 |
| CONTRAST TO AGEISM | ageism | 4546 | 1557 |
| CONTRAST TO BULLYING | bullying | 24619 | 7772 |
| ANTI-RACISM | racism | 47045 | 21991 |
| FIGHTING BIPOC DISCRIMINATION | bipoc | 685 | 332 |
| INTERCULTURAL EDUCATION | "intercultural education" | 1679 | 928 |
| DIVERSITY, INCLUSION, EQUALITY | diversity AND inclusion AND equality | 909 | 449 |
| GENDER EQUITY | "gender equity" | 6231 | 2558 |
| ANTI-SEXISM | sexism | 12003 | 4229 |
| FIGHTING VIOLENCE AGAINST WOMEN | violence AND against AND women | 15369 | 6186 |
| FEMINISM | feminism | 32978 | 18092 |
| CIVIL RIGHTS | "civil rights" | 22525 | 8216 |

| TOPIC | KEYWORD | ALL | FILTERS |
|---|---|---|---|
| HUMAN RIGHTS | "human rights" | 128288 | 48502 |
| EQUAL OPPORTUNITIES | "equal opportunit*" | 7024 | 3531 |
| SOCIAL JUSTICE | "social justice" | 48588 | 23197 |
| MINORITY RIGHTS | "minority right*" | 1860 | 1130 |
| INTERGENERATIONAL JUSTICE | intergenerational AND justice | 1314 | 630 |

A sample of 366 articles was thus obtained which are those reported in the References section.

The selected articles were then analyzed to identify the characteristics and dimensions of interest for the construction of the questionnaire.

The analysis of scientific articles published on various topics of interest therefore led to the creation of a conceptual, with the aim of providing an all-encompassing vision divided into Macro Areas of all the parameters of interest. The dimensions of the framework were then converted into component questions of the questionnaire.

The use of a systematic review of the scientific literature has therefore guaranteed as complete a vision as possible on the topics of interest, with the aim of making explicit and comprehensible the intrinsic complexity that characterizes sustainability, and particularly the social dimension of the latter.

**Questionnaire structure**

The framework resulting from the literature review and analysis led to the development of a questionnaire named S-Assessment tool for Social Sustainability.

The "S" Assessment is available, and it has been already tested on a benchmark of companies.[5]

This assessment tool is structured in five sections, corresponding to the SDGs30 and associated values of the ReWriters manifesto. The evaluation criteria have been determined to allow the identification of best practices carried on by sustainability-leading organizations, and to suggest development paths to become change agents, involving people in a sustainable transformation journey.

The questionnaire starts with an introductory section aimed at gathering descriptive information, i.e. number of employees, revenues, industry, societal form, and operating markets (B2C, B2B, B2G).

---

[5] Certificazione etica aziendale | ReWorld (re-world.it)

The first section, labelled as "Health and Wellness" gathers indicators pertaining to the dimensions of Bioethics Culture, Affective Education and Mental Health, and Body Positivity. This section contains a set of seven questions (plus an optional one) mainly aimed at evaluating whether the organization introduced specific measures concerning the above aspects, like for instance mental health support programs or contrast to appearance-based discriminations (e.g. Sullivan and Lewis, 2001; Scott et al., 2012; Schmidt and Cohn, 2021; Smith et al., 2023).

The second section, "Gender Equality", focuses on Gender equity, Anti-sexism, Fighting violence against women, Feminisms, together with Inclusive language, Non-violent communication, and Media ethics. These dimensions are measured through a set of sixteen questions (plus four optional). Like in the previous section, most questions act as a control tool for specific actions that organizations should undertake, like to support gender equality, women's rights, or contrasting the excessive use of male gender-based terms in Italian speech. Furthermore, this section also contains questions to evaluate the percentage of women at different hierarchical levels of the organization, and the distribution of average compensations for men and women at different hierarchical levels (e.g. Taylor and Walker, 1998; Tausig and Fenwick, 2001; Tatli et al., 2011; Seron et al., 2016; Smith et al., 2023).

The third section is named "Decent Work and Economic Growth", addressing key aspects that concern the following dimensions: Contrast to ableism, to ageism, to bullying, Collective intelligence and Cooperation as the key to development, Digital responsibility, and Work-Life balance. The set of fifteen questions (plus three optional) carries on the approach explained for the previous ones, evaluating whether the organization has put in place measures to support positive features and/or to contrast negative ones, and eventually which ones have been developed and adopted (e.g. Shields and Price, 2002; Spar and La Mure, 2003; Truxillo et al., 2015, Tombleson and Wolf, 2017).

The fourth section focuses on "Reducing inequalities" and gathers the higher number of indicators, about Youths and Intergenerational Justice, Civil and Human Rights, Minorities Rights, Equal opportunities and social justice, Diversity, Inclusion and Equality, Antiracism, Contrast to BIPOC discriminations, Intercultural education, Queerness and LGBTQI+ rights, Intersectionality (e.g. Tracy and Alberts, 2006; Trudel and Cotte, 2009; Triana et al., 2015; Tatli et al., 2017). Furthermore, it also considers aspects of other indicators that were partially addressed in previous sections, i.e. Contrast to ableism, to ageism, to bullying, Inclusive language, Non-violent communication, Media ethics. The evaluation approach for this section is the same exposed for the previous ones, structured in twenty-three questions (plus eleven optional).

The final section is labelled "Responsible Production and Consumption" and focuses on aspects and dimensions that are mostly related to Environmental Sustainability. Indeed, the six questions (plus five optional) are aimed at investigating how the organization promotes environmental sustainability between its employees and stakeholders, which environment-related politics are put in place, and if and how it supports animal rights, whether relevant for the specific business considered (e.g. Vuontisjärvi, 2006; Windsor, 2006; Vachhani and Pullen, 2019; Valente et al., 2020).

The answers provided in each section will act as inputs to conduct an evaluation of the company. The scores in each section are calculated as the sum of the individual points attributed to each answer; the score obtained is compared to the maximum achievable in each specific section.

Furthermore, some questions are subject to validation by researchers and the score is assigned only in case of a positive outcome. At the end of each section, companies can describe experiences and initiatives related to the objectives and upload supporting materials. These contributions directly impact the definition of the final score of the section in terms of a potential bonus to be achieved.

Progressive thresholds have been established which allow companies to be classified in four different categories, based on the score obtained, and to offer specific consultancy and training courses. These categories have been labelled as: Pioneer (score less than or equal to 30% of the maximum achievable), Builder (score between 31% and 60%), Transformer (score between 61% and 80%), Leader (with a score greater than 81%).

**Conclusions**

Through a meticulous methodology combining systematic literature review and narrative analysis, we have developed the S-Assessment Tool for Social Sustainability. This tool, structured around the dimensions of the ReWriters manifesto and the SDGs, serves as a robust framework for evaluating organizations' performance in the realm of social sustainability.

By providing organizations with a practical and comprehensive questionnaire, we aim to empower them to better understand and address their social impact. Through the identification of best practices and development paths, organizations can become change agents, actively involving people in a sustainable transformation journey.

The structured questionnaire encompasses a wide array of dimensions, including health and wellness, gender equality, decent work and economic growth, reducing inequalities, and responsible production and consumption. By addressing these dimensions, organizations can promote inclusivity, diversity, and equality within their operations and communities.

Through the implementation of the S-Assessment Tool, organizations can contribute to the achievement of the SDGs outlined in the 2030 Agenda. By aligning their practices with the values of the ReWriters manifesto and the principles of social sustainability, organizations can play a vital role in fostering a more just, equitable, and sustainable world for all.

As we move forward, it is essential to continue refining and validating the S-Assessment Tool, incorporating feedback from organizations and stakeholders, and adapting it to evolving social and environmental challenges. Additionally, further research could explore the impact of organizations' social sustainability efforts on their financial performance, stakeholder engagement, and overall resilience in the face of global challenges.

Overall, the development of the S-Assessment Tool represents a significant step towards promoting social sustainability and advancing the collective effort towards a more equitable and sustainable future for all stakeholders.